%% file: main.tex
\documentstyle[aps,preprint,prl,citesort,epsf]{revtex}

\newcommand{\eqref}[1]{Eq.~(\protect\ref{#1})}
\newcommand{\figref}[1]{Fig.~\protect\ref{#1}}

\begin{document}

\draft

\input{title-and-abstract}

\narrowtext

\input{body}

\input{acknowledgements}


\bibliographystyle{prsty}  



\input{references}
\newpage
\input{figures}

\centerline{\epsfysize=8.5in \epsfbox{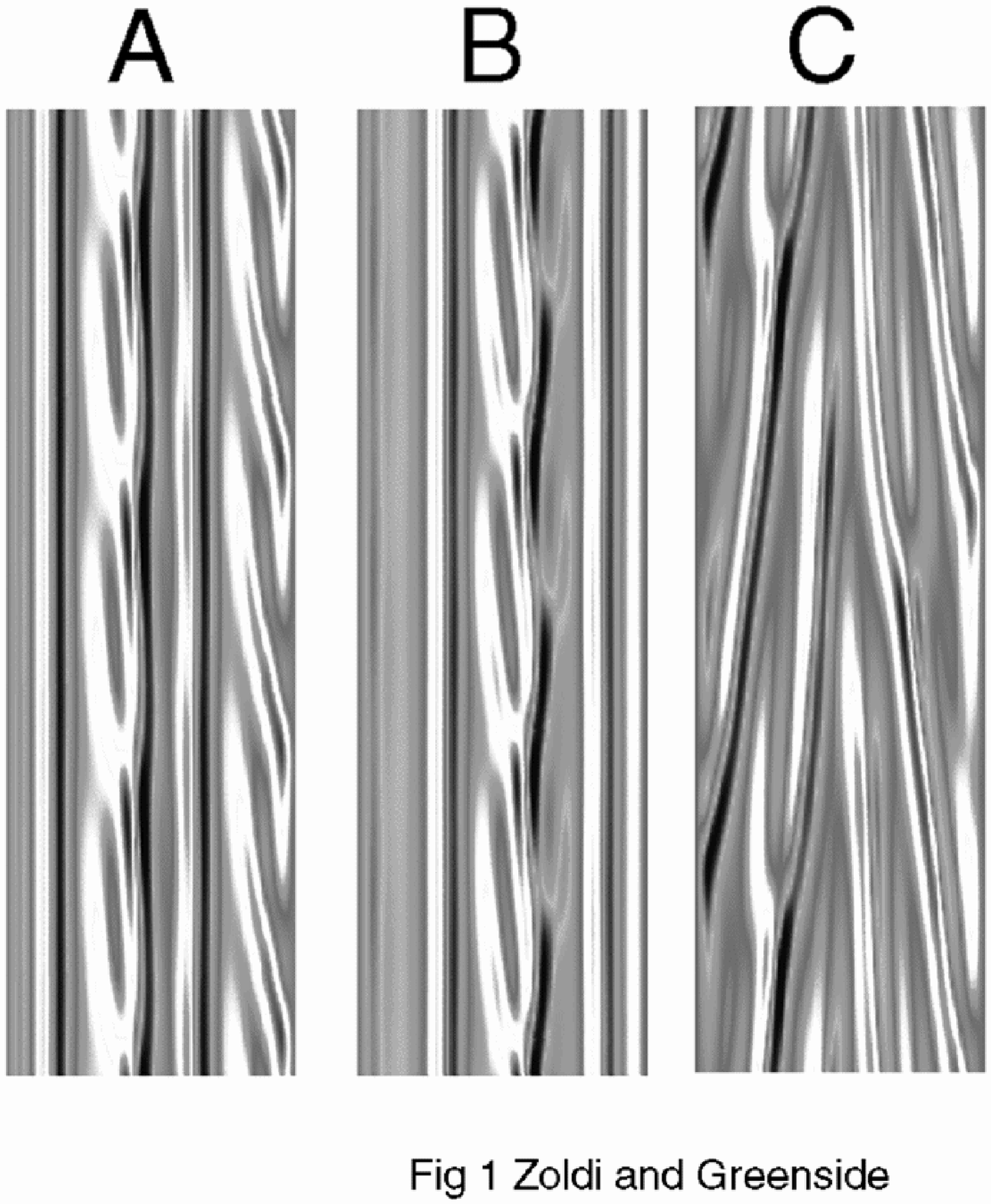}}
\centerline{\epsfysize=8.5in \epsfbox{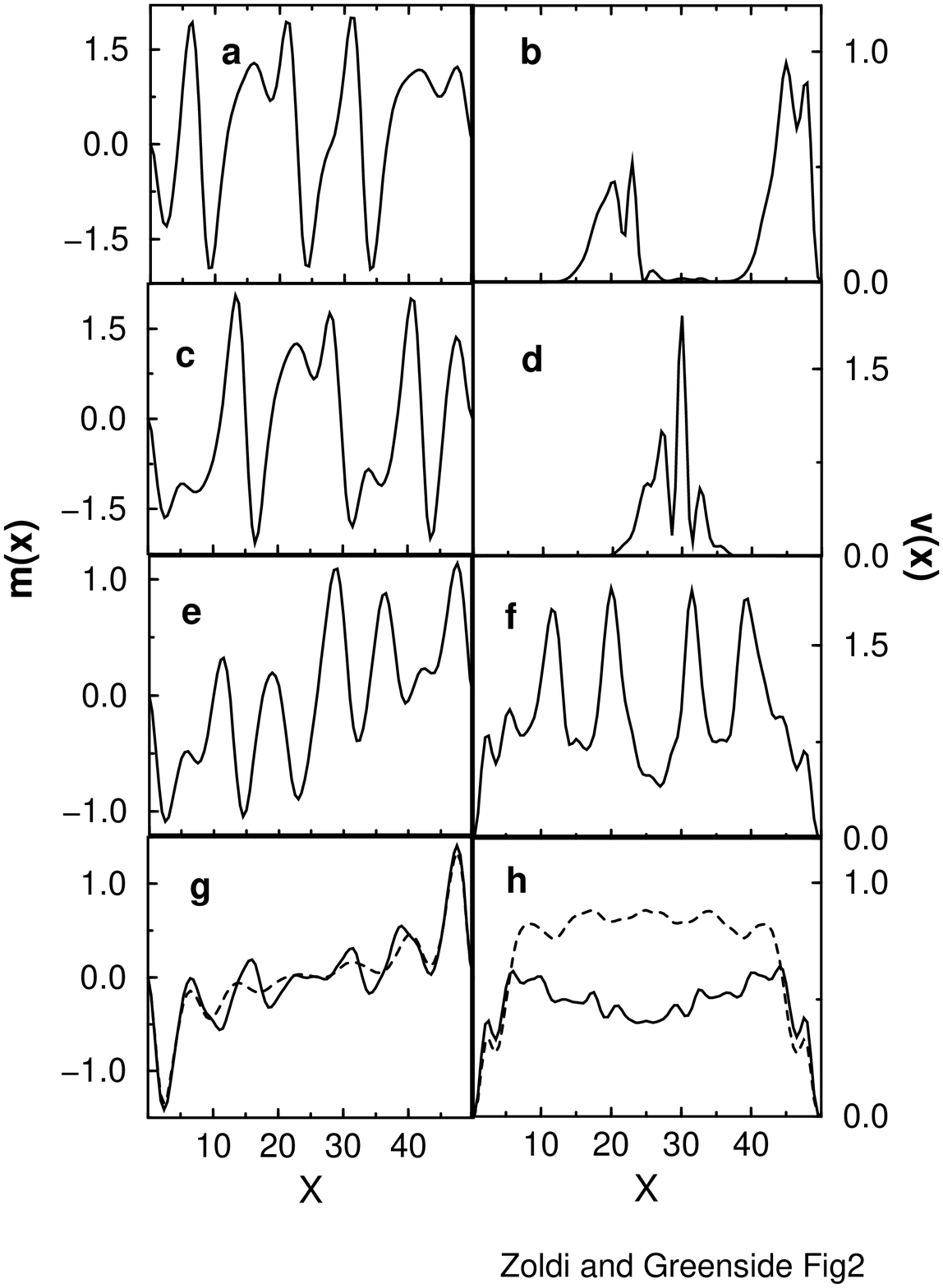}}
\centerline{\epsfysize=8.5in \epsfbox{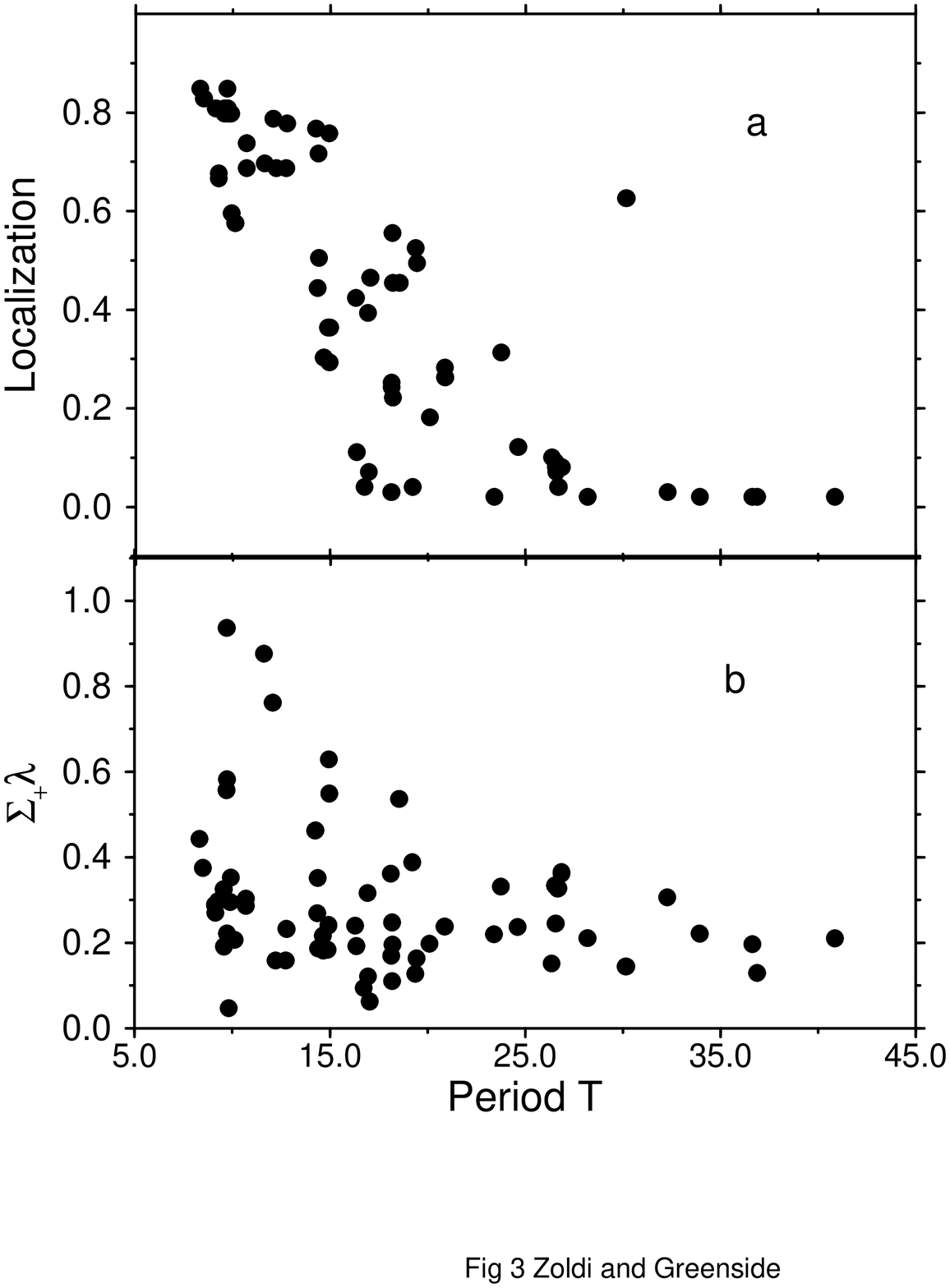}}

\end{document}

%% file: title-and-abstract.tex
\title{Spatially Localized Unstable Periodic Orbits}

\author{
Scott~M.\ Zoldi\cite{CNCS-address}\cite{zoldi-email} and
Henry~S.\ Greenside\cite{CNCS-address}
}

\address{
Department of Physics\\
Duke University, Durham, NC 27708-0305
}

\date{April 2, 1997}

\maketitle


\begin{abstract}
Using an innovative damped-Newton method, we report the first calculation
of many distinct unstable periodic orbits (UPOs) of a large
high-dimensional extensively chaotic partial differential equation. A
majority of the UPOs turn out to be spatially localized in that time
dependence occurs only on portions of the spatial domain. With a particular
weighting of 127~UPOs, the Lyapunov fractal dimension~$D=8.8$ can be
estimated with a relative error of~2\%. We discuss the implications of
these spatially localized UPOs for understanding and controlling
spatiotemporal chaos.
\end{abstract}


%% file: body.tex
Much recent research on sustained nonequilibrium systems has concerned
using unstable periodic orbits (UPOs) to characterize and to control chaos
\cite{Artuso90I,Politirefs,Badii94,Pierson95,Ding96}. The starting
point for this research has been the observation that a dense set of
unstable periodic orbits (UPOs) is associated with a strange attractor.
\cite{Guckenheimer83}. For low-dimensional chaos with a symbolic dynamics
(a unique labeling of the UPOs in the dense set
\cite{Guckenheimer83}), researchers have shown that
knowledge of a small number of UPOs can be used to improve forecasting
\cite{Pawelzik91} and to estimate dynamical invariants of the strange
attractor such as its fractal dimensions and Lyapunov
exponents~\cite{Artuso90I}. In some low-dimensional chaotic experimental
systems, enough UPOs have been determined from empirical data to apply this
formalism successfully~\cite{Badii94}. An exciting related advance has been
the discovery that these low-dimensional chaotic systems can be controlled
by stabilizing particular members of the dense set of UPOs with weak
time-dependent perturbations of a system parameter~\cite{control-refs}.

An important question is whether the above results
generalize and remain useful for high-dimensional dynamical
systems, especially for large nonequilibrium systems that
are extensively chaotic \cite{Cross93,Greenside97Montreal}.
We are particularly interested in two potential
applications.  One is to characterize extensive chaos in
terms of some finite number of UPOs \cite{Politirefs}. Large nonequilibrium
chaotic systems are difficult to analyze because of
long transient times (which, in some cases, increase
exponentially with system size \cite{supertransients}) and
also because it is often not known in advance how long a
system should be observed in a non-transient regime to
obtain good statistics. An attractive feature of UPOs
associated with a strange attractor is that {\em they can be
computed within a finite time interval} and so their
analysis replaces difficult questions of transients and
observation times with questions of mathematical and
numerical convergence, e.g., the number of UPOs needed to
approximate the invariant measure of an attractor to some
desired accuracy and whether sufficiently many UPOs can be
calculated numerically.  A second potential application is
to improve control of extensively chaotic systems. If the
magnitude of a control perturbation is to remain small as a
chaotic system becomes large, the system parameter will need
to be varied at spatially distributed control points
\cite{distributed-control}. An improved understanding of the
spatial structure of UPOs, of the distribution of their
periods~$T$, and of their stability should aid the
development of high-dimensional spatiotemporal control
algorithms by suggesting the number and location of control
points for a particular UPO and for a particular system
parameter.

In this Letter, we take a significant step towards
understanding the relation of the dense set of UPOs to
high-dimensional spatiotemporal chaos by reporting the first
calculation of many (over 100) distinct UPOs for a
high-dimensional ($D= 8.8$) driven-dissipative system partial
differential equation (pde) in an
extensively chaotic regime \cite{comment-on-Politi}. This calculation 
represents two
achievements. One is numerical, that a simple modification
of a Newton algorithm by the addition of damping
\cite{Dennis83} greatly increases the likelihood of
convergence and so makes
practical the computation of many UPOs from the nonlinear
equations that they satisfy. The second achievement is
several discoveries in nonequilibrium physics made possible
by this numerical method: that over 100~distinct UPOs can be
calculated in a large chaotic pde and therefore their
properties can be studied for the first time; that most of
these UPOs turn out to be spatially localized (as discussed
below); that about 100~UPOs are already sufficient to
estimate the fractal dimension of the high-dimensional
attractor to two significant digits; and that certain mean
quantities such as a time-averaged spatial pattern can be
estimated from a knowledge of the UPOs.

Especially interesting is the result that most of the UPOs
have dynamics that are spatially localized in that the
variance of temporal fluctuations is substantial only on
portions of the spatial domain (see
Figs.~\ref{fig:upo-pictures} and~\ref{fig:upo-details}
below). We speculate that this will be a general feature of
UPOs for extensively chaotic systems. This localization, in
fact, helps to explain how the spatiotemporal disorder of an
extensively chaotic system (which can be interpreted as many
small dynamically-independent subsystems
\cite{Greenside97Montreal}) can be consistent with the
global spatiotemporal coherence of each UPO in the dense
set. The temporal disorder can be understood as an irregular
wandering of a chaotic orbit near many UPOs (the case for
low-dimensional chaos) while the spatial disorder can now be
better understood as a complicated weighting of localized
spatially-irregular dynamics associated with the UPOs. The
spatially localized dynamics also has implications for
distributed control algorithms since control points placed
in the weakly time-independent spatial regions of a
particular UPO will likely not aid in stabilizing that
UPO. It is then important to know the spatial structure of
various UPOs before attempting to control them.

Our calculations were carried out for one of the simplest
models of extensive chaos, the one-dimensional
Kuramoto-Sivashinsky (KS) equation \cite{Cross93}
\begin{equation}
  \partial_t u  =   -  u \, \partial_x u
                    - \partial_x^2 u
                    -  \partial_x^4 u
  , \qquad x \in [0, L]
  , \label{ks-eq}
\end{equation}
for a field~$u(t,x)$ that lives on an interval of length~$L$
and that satisfies ``rigid'' boundary conditions $u
=\partial_x{u} = 0$ at~$x=0$ and~$x=L$.  For system sizes~$L
\ge 50$, Manneville has shown that typical initial
conditions evolve towards a chaotic attractor that is
extensive in that the Lyapunov fractal dimension~$D$
increases linearly with~$L$ \cite{Manneville85}. In our
calculations, we chose a fixed length~$L=50$ and spatial
resolution~$\triangle{x}=0.5$ for which the Lyapunov fractal
dimension was~$D=8.8$ and there were 4~positive, 1~zero, and
94~negative Lyapunov exponents \cite{Christiansen-comment}.
The system size~$L=50$ was just large enough to be in the
extensively chaotic regime and yet small enough that the
numerical calculations were manageable with available
resources and algorithms. The reason for emphasizing the
extensively chaotic regime was that certain details of the
spatiotemporal chaos and of the UPOs could then be expected
to be insensitive to the value of the system length~$L$.

UPOs of \eqref{ks-eq} were calculated numerically by using a
damped-Newton method \cite{Dennis83} with simple shooting
\cite[pages 120-122]{Parker89} to solve a set of
nonlinear equations ${\bf F}({\bf X})={\bf 0}$ for unknowns
${\bf X} = (T,{\bf U}_0)$ which we define as follows.  Given
a uniformly spaced mesh~$x_i = i \triangle{x}$ on the
interval~$[0,L]$ with spatial resolution
$\triangle{x}=L/(N+1)$, we denote the field value at
the~$i$th mesh point at time~$t$ by $u_i(t)=u(x_i,t)$ for
$i=0,\cdots,N+1$. Then the $(N+1)$-dimensional vector of
unknowns~$\bf X$ consists of the UPO's period~$T$ and of a
$N$-dimensional point~${\bf U}_0 = (u_1(0),\cdots,u_N(0))$
that lies on the UPO. The values $u_0=u(t,0)$
and~$u_{N+1}=u(t,L)$ are zero by the boundary
conditions and do not need to be solved for.

The corresponding $(N+1)$-dimensional nonlinear
equations~${\bf F}({\bf X})=0$ for~$\bf X$ are determined by
trying to find a closed orbit~${\bf U}_t$ of duration~$T$
starting from~${\bf U}_0$, i.e., such that~${\bf U}_T - {\bf
U}_0 = {\bf 0}$. These nonlinear equations are derived and
solved as follows. First, \eqref{ks-eq} is reduced to a
$N$-dimensional set of first-order autonomous odes,
\begin{equation}
  d{\bf U}/dt = {\bf G}({\bf U}) ,
  \qquad {\bf U}(0) = {\bf U}_0
  . \label{discrete-ks-eq}
\end{equation}
by replacing the spatial derivatives in
\eqref{ks-eq} and in the boundary condition
$\partial_x{u}=0$ with second-order accurate
finite-difference approximations. The vector~${\bf U}_T$ at
time~$T$ is then found by numerical integration of
\eqref{discrete-ks-eq}. Second, a $N\times N$
matrix~${\bf M}_T$ at time~$T$ is obtained by integrating
the $N^2$ linear variational equations \cite{Parker89}
\begin{equation}
  d{\bf M}/dt = {\bf J} {\bf M} ,
  \qquad {\bf M}(0) = {\bf I}
  , \label{variational-eqs}
\end{equation}
together with \eqref{discrete-ks-eq}, where ${\bf J}(t) =
\partial{\bf G} / \partial{\bf U}$ denotes the $N \times N$
Jacobian matrix and $\bf I$ denotes the~$N \times N$
identity matrix. (The matrix~${\bf M}(t)$ is the propagator
that evolves infinitesimal perturbations from time~$t=0$ to
time~$t$.)  Given the vector~${\bf U}_T$ and matrix~${\bf
M}_T$, a Newton correction $\delta{\bf X} =
(\delta{T},\delta{\bf U})$ for updating the vector~$\bf X$
is found by solving the following $2 \times 2$~block matrix
equation~\cite{Parker89}:
\begin{equation}
  \left(
    \begin{array}{cc}
      {\bf 0}  &  {\bf G}({\bf U}_0)^\dagger  \\
      {\bf G}({\bf U}_T)  &  {\bf M}_T - {\bf I}
    \end{array}
  \right)
  \left(
    \begin{array}{c}
      \delta{T} \\
      \delta{\bf U}
    \end{array}
  \right)
  =
  \left(
    \begin{array}{c}
      {\bf 0} \\
      {\bf U}_0- {\bf U}_T
    \end{array}
  \right)
  , \label{block-Newton-eq}
\end{equation}
where the symbol~$\dagger$ denotes the matrix transpose.  A
single Newton step is then performed by updating the present
values for the unknowns, ${\bf X} \to {\bf X} +\delta{\bf
X}$, and such steps are repeated until the residuals and
corrections are sufficiently small
\begin{equation}
  \|{\bf U}_T - {\bf U}_0\|_\infty < 10^{-3} \|{\bf U}_0\|_\infty ,
  \qquad \mbox{and} \qquad
  \|\delta{\bf X}\|_\infty < 10^{-3} \| {\bf X}_0 \|_\infty
  , \label{Newton-convergence-criteria}
\end{equation}
in the infinity norm $\|{\bf X}\|_\infty = \max_i
|X_i|$. Each Newton step requires the time-integration of a
set of~$N + N^2$ odes over time~$T$ plus the solution of the
linear equations \eqref{block-Newton-eq}.  For the
calculations reported below, we used time-splitting methods
for Eqs.~(\ref{discrete-ks-eq}) and~(\ref{variational-eqs})
that were first-order accurate in time and second-order
accurate in space with a constant time step
$\triangle{t}=0.005$. Since the block matrix ${\bf M}_T -
{\bf I}$ in \eqref{block-Newton-eq} is typically dense,
$PLU$-factorization was used to solve for the correction;
the time to carry out the linear algebra was much larger the
time to integrate Eqs.~(\ref{discrete-ks-eq})
and~(\ref{variational-eqs}) over a period~$T$. Most UPOs
that we calculated were robust to modest changes in
spatiotemporal resolution and the Newton convergence
criteria.

For high-dimensional Newton methods, it is essential to have a good
starting guess~${\bf X}_0$ since Newton methods are guaranteed to converge
only locally. We initially tried to find a good initial guess $(T,{\bf
U}_0)$ by searching for approximate recurrences \cite{Badii94,Pawelzik91}
of chaotic time series~${\bf U}_i = {\bf U}(i \triangle{t})$ in the
high-dimensional numerical phase space of
\eqref{discrete-ks-eq}. This turned out to be impractical,
e.g., for~$L=50$, integration times of at least~$10^8$ time units were
needed to find a single approximate recurrence of period~$T=11.6$ 
within a ball of rather
large radius~$0.1$, $\|{\bf U}_T - {\bf U}_0\|_\infty < 0.1$
\cite{warning-for-control}. Further,  these approximate
recurrences did not turn out to be close to any of the UPOs that we calculated
using the more rigorous Newton method \cite{distinct-criteria}.  
Since these results suggest that no
approximate recurrence is close to a UPO of \eqref{ks-eq}, we then tried to
choose an initial state by choosing a positive random number~$T$ for the
period and an initial vector~${\bf U}_0$ from a point on the chaotic
attractor. Using this procedure, the convergence criteria
\eqref{Newton-convergence-criteria} always failed within the
specified maximum number of~200 Newton iterations.

Convergence was finally obtained with a {\em damped}-Newton method
\cite{Dennis83}, in which only a fraction~$\alpha \le 1$ of a Newton
correction~$\delta{\bf X}$ was added to update the unknowns, ${\bf X} \to
{\bf X} +
\alpha \delta{\bf X}$. We used a particular damping method
known as the Armijo rule \cite{Dennis83} which expresses the damping
factor in the form~$\alpha = 2^{-j}$.  During each Newton iteration,
successive values of the integer $j=0,1,\cdots$ were tested until a
value~$j$ was found such that the new residual $\|{\bf F}({\bf X} +
2^{-j}\delta{\bf X})\|$ was smaller by a factor $1 -
\alpha/2$ than the previous residual $\|{\bf F}({\bf
X})\|$ in the Euclidean norm. With the Armijo rule, Newton's method
converged with the criteria
\eqref{Newton-convergence-criteria} for roughly five percent
of all initial guesses~$(T,{\bf U}_0)$ consisting of a positive random
number~$T$ and some point~${\bf U}_0$ on the chaotic attractor.
If the damping parameter~$\alpha$ fell below the value~$2^{-6}$, it was
more efficient to terminate the Newton iteration and start over with a new
guess~$(T,{\bf U}_0)$.  The specific value of~$j=6$ was chosen as a trade
off between convergence and computer time, as tiny fractions of the
correction can lead to an excessive amount of time spent on a single guess.

We now discuss the properties of the UPOs calculated with the above
numerical methods.  Using the Armijo rule and the Newton method discussed
above, 127~distinct UPOs were found from 5000 guess~${\bf X}_0 =
(T_0,{\bf U}_0)$ \cite{distinct-criteria}.  UPOs with periods shorter
than~$8$ could not be found while the Armijo-Newton algorithm failed to
converge for UPOs with periods larger than 42, probably because the simple
shooting method becomes unstable for periods larger than about this value.

As shown qualitatively in \figref{fig:upo-pictures} and more
quantitatively in \figref{fig:upo-details}, most of the UPOs
are localized in that their time variation is substantial
only in isolated portions of the domain.
\figref{fig:upo-pictures}(A) shows a state for which the
temporal variation is localized in two small spatial
intervals, one of which is adjacent to the right
boundary. \figref{fig:upo-pictures}(B) shows an interior
state whose dynamics is localized to a single interval well
away from the boundaries. \figref{fig:upo-pictures}(C) shows
an extended UPO in which the temporal variation occurs
throughout the spatial domain. The corresponding mean and
variance patterns \cite{distinct-criteria} are given
in \figref{fig:upo-details}(a)-(f). In
\figref{fig:upo-details}(a), (c), and~(e), the mean pattern
is nonzero throughout the domain which holds for the other
124 UPOs as well.
\figref{fig:upo-details}(b) and~(d) indicate more clearly
the localization of the dynamics, which is evidently
uncorrelated with the mean pattern. The variance decreases
three to five orders of magnitude outside the regions of
substantial variation.

From an explicit knowledge of the space-time evolution of the 127~distinct
UPOs, the time-averaged mean and the variance of the spatiotemporal
chaotic solution of KS equation \eqref{ks-eq} can be estimated. A simple 
averaging~$<m(x)>_{\rm UPO}$ of all 127~mean patterns $m(x)$ for the UPOs yields the solid 
curve in \figref{fig:upo-details}(g), which should be compared with the 
dashed curve obtained by averaging an extensively chaotic field~$u(t,x)$ 
over $10^6$ time units.  The relative error in the infinity norm 
between the patterns is 24\% and is substantially better near the boundaries.  Strikingly, 
$<m(x)>_{\rm UPO}$ has the same qualitative structure as the 
mean pattern of the 
attractor $<u(x)>$ using only a moderate number of low-period UPOs. 
In contrast, an average of the 127~variance
patterns (solid line in \figref{fig:upo-details}(h)) does not agree as
well (a relative error of~$46\%$) with the variance of the 
extensively chaotic
field~$u(t,x)$ averaged over~$10^{6}$ time units but still reflects some
of the qualitative features. These results suggest that
a knowledge of UPOs may help to understand the interesting mean patterns
and localized dynamics observed recently in Faraday wave experiments
\cite{crispation-expts}.

\figref{fig:instability-and-fraction-vs-period} shows how the extent of
localization and instability depends on the period~$T$.
To characterize the degree of localization of the UPOs, a localization
number was defined as the fraction of the interval~$[0,L]$ for which the
variance~$v(x)$ was smaller than~$0.05$ (the results were not sensitive to
the choice of this cutoff).  
\figref{fig:instability-and-fraction-vs-period}(a) shows the
localization number versus the period~$T$ for all 127~UPOs.  Although there
is scatter in the points, we can draw a few conclusions: that UPOs with
period less than about 8 don't exist; that shorter period UPOs are more
strongly localized; and that there can be many UPOs of approximately the
same period (say~$T=14$) and these can vary substantially in their
localization.  The outlier point with a localization of~$0.6$ for
period~$T=30$ suggests that the right side of this curve is incomplete,
that there are localized UPOs of higher period but our numerical algorithms
were not able to find them. In 
\figref{fig:instability-and-fraction-vs-period}(b), we summarize the
instability of all 127~UPOs as a function of
their period~$T$. The instability of a UPO is defined as the sum $\sum_{+}
\lambda$ of all positive transverse Lyapunov exponents which are defined
by~$\lambda =
\log(|m|)/T$, where~$m$ is a Floquet multiplier, i.e.,
an eigenvalue of the propagator matrix~${\bf M}_T$ in
\eqref{variational-eqs} evaluated for a UPO
of period~$T$ \cite{Parker89}.
\figref{fig:instability-and-fraction-vs-period}(b)
shows that, on average, smaller period UPOs are more unstable. Again the
right side of this curve is incomplete and it would be interesting to
extend the data using more sophisticated numerical algorithms.

Using the data in \figref{fig:instability-and-fraction-vs-period}(b) one
can estimate the fractal dimension~$D$ of the chaotic attractor. First, a
fractal dimension is formerly associated with each UPO by generalizing the
Kaplan-Yorke formula \cite{Parker89} to UPOs in terms of their transverse
Lyapunov exponents; we find dimensions ranging from 6 to~12 for the
127~UPOs. 
The fractal dimension of
the chaotic attractor can be viewed as an average over the fractal
dimensions of the UPOs which are associated with the attractor.  To obtain
the fractal dimension over attractor, each UPO's dimension is summed with a
weighting $1/(\sum_+\lambda)$ (less unstable UPOs are weighted more heavily)
which gives an estimate of~$D=9.0$ to the fractal dimension of the
attractor. This is in good agreement with the Lyapunov fractal dimension
$D=8.8 \pm 0.1$ calculated directly from the Lyapunov exponents of the
spatiotemporal chaotic solution of \eqref{ks-eq} \cite{Manneville85}. The
convergence to the Lyapunov dimension of the estimate based on UPOs is
statistical in that the error decreases approximately as~$1/\sqrt{N}$
where~$N$ is the number of UPOs contributing to the weighted sum. Other
weightings of the UPOs were tried
\cite{Pawelzik91,Grebogi88} but found not to give results as satisfactory
as the $1/(\sum\lambda_+)$ weighting.

In conclusion, we have used an innovative numerical method to calculate for
the first time many UPOs associated with a large high-dimensional pde, \eqref{ks-eq}.  An important numerical insight was the use of
damping to increase the likelihood of convergence of an otherwise
straightforward Newton method.  Near-recurrences in the time-series data as
long as $10^{8}$ time units were found not to correspond to any computed
UPOs.  From the 127~distinct UPOs found using the damped-Newton method, we
could estimate the fractal dimension of the attractor to be~$9.0
\pm 0.1$ compared to the actual value of~$8.8 \pm 0.1$.  These UPOs were
also used to predict successfully the qualitative features of the
time-averaged mean pattern and the variance of the chaotic attractor.  A
new insight is that the UPOs are typically localized in space.  This
localization suggests a new way to think about the dynamically independent
subsystems within extensive chaos.  The localization also has important
implications for the control of large chaotic systems using distributed
sets of control points.

The present calculations raise two interesting questions for
future work. One is to improve the numerical algorithms,
primarily by using multishooting methods \cite{Ascher95}, so
that UPOs of partial differential equations in two- and
three-space dimensions can be calculated. It will then be
quite interesting to determine whether there are systematic
patterns of defects associated with UPOs as their period
increases from their smallest value. The second question is
to understand the implications of localized UPOs for
spatially distributed control, e.g., by the OGY or
delayed-feedback methods.

%% file: acknowledgements.tex
We thank L.~Bunimovich, S.~Newhouse and M.~Strain for useful
discussions.  This work was supported by a DOE Computational
Science Graduate Fellowship, by NSF grants NSF-DMS-93-07893
and NSF-CDA-92123483-04, and by DOE grant
DOE-DE-FG05-94ER25214.

%% file: figures.tex
\begin{figure}   
\caption{ Density plots of three representative UPOs
$u(t,x)$ calculated by applying a damped-Newton method to
the Kuramoto-Sivashinsky equation \eqref{ks-eq} in a spatial
domain of length~$L=50$.  The horizontal axis is space and
the vertical axis spans a time interval of 35~time units.  {\bf
(a)} A UPO of period~$T=9.9$ with dynamics localized near
the right boundary; {\bf (b)} A UPO of period~$T=10.7$ with
dynamics localized in interior of the interval; {\bf (c)} An
extended UPO of period~$T=23.4$. The greyscales represent
amplitude variations between about 3 and -3. }
\label{fig:upo-pictures}
\end{figure}

\begin{figure}   
\caption{Time-averaged mean patterns~$m(x)=<u(x,t)>$ and
variance patterns~$v(x) = <(u(x,t) - m(x))^2>$ for the three
representative UPOs of \figref{fig:upo-pictures}. {\bf (a)}
and {\bf (b)}: for the UPO with dynamics localized near a
boundary. {\bf (c)} and {\bf (d)}: for the UPO with dynamics
localized away from boundaries.  {\bf (e)} and {\bf (f)}:
for the extended UPO.  {\bf (g)} and {\bf (h)}: mean and
variance patterns (solid lines) averaged over all~127
distinct UPOs. For comparison, the dashed lines give the
corresponding mean and variance patterns obtained from an
integration of a chaotic solution over~$10^6$ time
units. }
\label{fig:upo-details}
\end{figure}

\begin{figure}   
\caption{ {\bf (a)}
Localization (fraction of the spatial domain that has variance~$v(x)$
below~$0.05$) versus period~$T$ of~127 distinct UPOs
calculated for the KS equation (\eqref{ks-eq}) in an
extensively chaotic regime with system size~$L=50$. 
{\bf (b)} Degree of instability as measured by
the sum~$\sum_+\lambda$ of positive transverse Lyapunov
exponents versus the period~$T$ for all 127~UPOs. }
\label{fig:instability-and-fraction-vs-period} 
\end{figure}

%% file: main.bbl
\begin{references}

\bibitem[*]{CNCS-address} Also Center for Nonlinear and
Complex Systems, Duke U., Durham, NC

\bibitem[\dagger]{zoldi-email} E-mail: zoldi@phy.duke.edu

\bibitem{Artuso90I}
R. Artuso, E. Aurell, and P. Cvitanovi\'c, Nonlinearity {\bf 3},  325  (1990).

\bibitem{Politirefs} A. Politi and A. Torcini,
Phys. Rev. Lett. {\bf 69}, 3421 (1992); A. Politi and
A. Torcini, Chaos {\bf 2}, (293) (1992).

\bibitem{Badii94}
R. Badii {\it et~al.}, Rev. Mod. Phys. {\bf 66},  1389  (1994).

\bibitem{Pierson95}
D. Pierson and F. Moss, Phys. Rev. Lett. {\bf 75},  2124  (1995).

\bibitem{Ding96}
M. Ding {\it et~al.}, Phys. Rev. E {\bf 53},  4334  (1996).

\bibitem{Guckenheimer83} J. Guckenheimer and P. Holmes, {\em
Nonlinear Oscillations, Dynamical Systems, and Bifurcations
of Vector Fields} (Springer-Verlag, New York, 1983).

\bibitem{Pawelzik91}
K. Pawelzik and H.~G. Schuster, Phys. Rev. A {\bf 43},  1808  (1991).

\bibitem{control-refs} E.~Ott, C.~Grebogi, and J.~A.\ Yorke,
Phys. Rev. Lett. {\bf 64}, 1196 (1990); K. Pyragas,
Phys. Lett. A {\bf 170}, 421 (1992); G.~Chen and X.~Dong,
Int.\ J.\ Bifurcation and Chaos {\bf 3}, 1363 (1993).

\bibitem{Cross93}
M.~C. Cross and P.~C. Hohenberg, Rev. Mod. Phys. {\bf 65},  851  (1993).

\bibitem{Greenside97Montreal}
H.~S. Greenside,  in {\em Semi-Analytic Methods for the Navier Stokes
  Equations}, {\em Proceedings of the CRM Workshop}, edited by K. Coughlin
  (Centre de Recherches Mathematiques, Montreal, Canada, 1997), in press; LANL
  chao-dyn/9612004.

\bibitem{supertransients} J.~P. Crutchfield and K. Kaneko,
Phys. Rev. Lett. {\bf 60}, 2715 (1988); Y.-C. Lai and
R.~L. Winslow, Phys. Rev. Lett. {\bf 74}, 5208 (1995).

\bibitem{distributed-control} I. Aranson, H. Levine, and
L. Tsimring, Phys. Rev. Lett. {\bf 72}, 2561 (1994);
D. Auerbach and J.~A. Yorke, J. Opt. Soc. Am. B {\bf 13},
2178 (1996); M.~E. Bleich, D. Hochheiser, J.~V. Moloney, and
J.~E.~S. Socolar, Phys. Rev. E, {\bf 55} (3) 2119 (1997).

\bibitem{comment-on-Politi} In previous work, Politi et al
\cite{Politirefs} have studied UPOs of a 1d extensively
chaotic system with {\em discrete} time, consisting of up to
six weakly coupled H\'enon maps. In contrast, our analysis
concerns a continuous-time strongly-coupled system with an
emphasis on the spatial structure of the UPOs.

\bibitem{Dennis83}
J.~E. Dennis and R. Schnabel, {\em Numerical Methods for Unconstrained
  Optimization and Nonlinear Equations} (Prentice-Hall, Englewood Cliffs, NJ,
  1983).

\bibitem{Manneville85}
P. Manneville,  in {\em Macroscopic modeling of turbulent flows}, Vol.~230 of
  {\em Lecture Notes in Physics}, edited by O. Pironneau (Springer-Verlag, New
  York, 1985), pp.\ 319--326.

\bibitem{Christiansen-comment} Recently, F. Christiansen,
P. Cvitanovi\'c and V. Putkaradze [Nonlinearity {\bf 10} 55
(1997)] also studied the KS~equation to investigate how a
chaotic attractor may be related to the UPOs whose closure
defines that attractor. Their emphasis and results differ
substantially from ours in that these authors considered a
smaller {\em periodic} domain
for which the chaotic dynamics is low-dimensional (we
estimate~$D \approx 2.2$ from their data) and for which the
set of UPOs has an explicit symbolic dynamics.

\bibitem{Parker89}
T.~S. Parker and L.~O. Chua, {\em Practical Numerical Algorithms for Chaotic
  Systems} (Springer-Verlag, New York, 1989).

\bibitem{warning-for-control} Our inability to find
approximate recurrences from numerical output~${\bf U}(t)$
suggests that some of the widely used control methods
\cite{control-refs} will not succeed in stabilizing UPOs
since they depend crucially on knowing an approximate
recurrence as input to the algorithm.

\bibitem{distinct-criteria} UPOs are distinct in that their
time-averaged pattern~$m(x)$ and variance~$v(x)$ averaged over one 
period,~$m(x)=\langle u(t,x) \rangle$ and~$v(x) = \left\langle (u(t,x) -
m(x))^2 \right\rangle$ differed by at least 0.01 and their periods differed by
at least 0.01.

\bibitem{crispation-expts} B.~J. Gluckman, C.~B. Arnold, and
J.~P. Gollub, Phys. Rev. E {\bf 51}, 1128 (1995);
A. Kudrolli and J.~P. Gollub, Physica D {\bf 97}, 133
(1996).

\bibitem{Grebogi88} C. Grebogi, E. Ott, and J.~A. Yorke,
Phys. Rev. A {\bf 37}, 1711 (1988).

\bibitem{Ascher95}
U.~M. Ascher, R.~M.~M. Mattheij, and R.~D. Russell, {\em Numerical Solution of
  Boundary Value Problems for Ordinary Differential Equations} (SIAM,
  Philadelphia, 1995).


\end{references}
